\documentclass[aps,prd,floats,twocolumn,epsf,nofootinbib]{revtex4}
\usepackage{latexsym}
\usepackage{amssymb}
\usepackage[dvips]{graphicx}
\usepackage[applemac]{inputenc}
\usepackage[nottoc]{tocbibind} 
\usepackage{float}
\usepackage{fancyhdr}
\usepackage{amsfonts}
\usepackage{amsmath}
\usepackage{color}
\usepackage{mathrsfs}
\usepackage{bm}
\usepackage{epsfig}

\newcommand{\be}{\begin{equation}}
\newcommand{\ee}{\end{equation}}

\def\lsim{\mathrel{\raise.3ex\hbox{$<$\kern-.75em\lower1ex\hbox{$\sim$}}}}

\def\gsim{\mathrel{\raise.3ex\hbox{$>$\kern-.75em\lower1ex\hbox{$\sim$}}}}

\def\beq{\begin{eqnarray}}
\def\eeq{\end{eqnarray}}
\def\bea{\begin{eqnarray}}
\def\eea{\end{eqnarray}}

\begin{document}

\title{A Consistent Dark Matter Interpretation For CoGeNT and
DAMA/LIBRA}

\author{Dan Hooper$^{a,b}$, J.I. Collar$^{c}$, Jeter Hall$^{a}$, Dan McKinsey$^{d}$ and Chris Kelso$^{e}$}

\address{
$^a$Center for Particle Astrophysics, Fermi National Accelerator
Laboratory, Batavia, IL 60510 \\ 
$^b$Department of Astronomy and Astrophysics, University of Chicago,
Chicago, IL 60637 \\
$^c$Enrico Fermi Institute, KICP and Department of Physics,
University of Chicago, Chicago, IL 60637 \\ 
$^d$Department of Physics, Yale University, New Haven, CT  06520 \\
$^e$Department of Physics, University of Chicago, Chicago, IL 60637}

\begin{abstract}

In this paper, we study the recent excess of low energy events
observed by the CoGeNT collaboration and the annual modulation
reported by the DAMA/LIBRA collaboration, and discuss whether these
signals could both be the result of the same elastically scattering
dark matter particle. We find that, without channeling but when
taking into account uncertainties in the relevant quenching factors,
a dark matter candidate with a mass of approximately $\sim$7.0 GeV
and a cross section with nucleons of $\sigma_{\rm DM-N}\sim 2 \times
10^{-4}$ pb ($2\times 10^{-40}$ cm$^2$) could account for both of
these observations. We also comment on the events recently observed
in the oxygen band of the CRESST experiment and point out that these
could potentially be explained by such a particle. Lastly, we compare
the region of parameter space favored by DAMA/LIBRA and CoGeNT to the
constraints from XENON 10, XENON 100, and CDMS (Si) and find that
these experiments cannot at this time rule out a dark matter
interpretation of these signals.

\end{abstract}

\maketitle


\section{Introduction}

For nearly a decade, the DAMA collaboration (and more recently, the
DAMA/LIBRA collaboration) has reported an annual modulation in their
event rate and interpreted this signal as evidence for particle dark
matter. According to their most recent results, which make use of
over 1.17 ton-years of data, the DAMA/LIBRA collaboration observes a
modulation with a significance of 8.9$\sigma$, and with a phase
consistent with that predicted for elastically scattering dark
matter~\cite{damanew}. When the null results from other dark matter
searches~\cite{cdms,xenon100} are taken into account, one is forced
to consider very light dark matter particles ($\lsim 10$ GeV) to
accommodate this signal~\cite{petriello}.\footnote{Alternatively, one
could also consider scenarios in which dark matter particles interact
with nuclei through a resonance~\cite{resonance}, interact with
nuclei with a momentum dependence causing them to scatter more
efficiently with NaI than other targets~\cite{ffdm}, or which
interact with nuclei largely through inelastic
processes~\cite{inelastic}; any of which could plausibly generate
the DAMA/LIBRA signal while evading all relevant null results.}

Recently, the CoGeNT collaboration has announced the observation of
an excess of low energy events relative to expected
backgrounds~\cite{cogentnew}. This excess, if interpreted as dark
matter, implies the dark matter particles possess a mass in the range
of 5-15 GeV and an elastic scattering cross section with nucleons on
the order of $10^{-4}$ pb ($10^{-40}$ cm$^2$). These implied values
are remarkably similar to those needed to generate the annual
modulation reported by the DAMA/LIBRA collaboration~\cite{theory}.

Dark matter interpretations of the combined DAMA/LIBRA and CoGeNT
signals have, however, been somewhat controversial. One reason for
this is that it has been claimed that the regions of dark matter
parameter space (mass vs.~cross section) implied by CoGeNT and
DAMA/LIBRA do not overlap, unless channeling occurs in
the DAMA/LIBRA
apparatus~\cite{cogentnew,compare1,compare2,compare3,compare4,compare5}. This
problem has been exacerbated by recent theoretical work which
suggests that the effects of channeling in DAMA/LIBRA should be
much smaller than previously considered~\cite{nochanneling} (even if 
some model-dependence remains). Another source of controversy has
resulted from the null results of other dark matter searches,
including XENON100, XENON10, and CDMS
(Si)~\cite{compare1,compare2,xenon100,Savage:2010tg}.

In this paper, we revisit these and related issues in an attempt to
determine whether the signals reported by the DAMA/LIBRA and CoGeNT
collaborations could potentially originate from the same dark matter
particle without conflicting with the null results of other
experiments. In Sec.~\ref{consistency}, we calculate the regions of
dark matter parameter space implied by DAMA/LIBRA and CoGeNT and
determine that, if uncertainties in these experiments' quenching
factors are taken into account, consistent regions do exist. In
particular, the combination of DAMA/LIBRA and CoGeNT data can be well
accommodated by a dark matter particle with a mass of approximately
$\sim$7 GeV and an elastic scattering cross section with nucleons of
$\sim 2 \times 10^{-4}$ pb ($2 \times 10^{-40}$ cm$^2$), even if no
significant channeling is taking place. We also comment on the events
recently observed in the oxygen band of the CRESST experiment. In
Sec.~\ref{null}, we discuss the null results of other dark matter
experiments, including XENON 10, XENON 100, and CDMS (Si), and find
that none currently exclude the region favored by the combination of
DAMA/LIBRA and CoGeNT. We summarize our result in
Sec.~\ref{conclusions}.



\section{Consistency of CoGeNT and DAMA/LIBRA}
\label{consistency}

Since the first presentation of the recent CoGeNT results four months
ago~\cite{cogentnew}, several
groups~\cite{compare1,compare2,compare3,compare4,compare5} have fit the
observed spectrum of events to elastically scattering dark matter
scenarios and compared these fits to those implied by the annual
modulation observed by DAMA/LIBRA~\cite{damanew}. While these studies
find that the CoGeNT and DAMA/LIBRA signals point to similar regions
of dark matter parameter space, the regions were found to overlap
only if the effects of channeling are significant within the
DAMA/LIBRA detectors. 

In channeled events, the crystal nature of the detector enables the
total recoil energy to be detected, in contrast to ordinary nuclear
recoil events in which only a fraction (known as the quenching
factor) of the energy is deposited in observable forms (scintillation
light, heat, and/or ionization) relative to that in electron
recoils~\cite{channeling,Bernabei:2007hw}. Recent theoretical work,
however, appears to disfavor the possibility that channeling plays an
important role in an experiment such as
DAMA/LIBRA~\cite{nochanneling,Savage:2010tg}. In particular, ions
recoiled by a dark matter particle originate in lattice sites and
will not approach the channels of the crystal, but instead are
expected to be efficiently blocked by the crystal lattice. In light
of these findings, we will assume throughout this study that the
fraction of events that are channeled at DAMA/LIBRA (or in other
direct detection experiments) is negligible.


The question we wish to address in this section is whether, without
channeling, the CoGeNT and DAMA/LIBRA signals could both originate
from the same dark matter particle species. With this goal in mind,
we consider the systematic uncertainties involved in these
experiments' results, in particular those pertaining to the germanium
and sodium quenching factors. 

Following Ref.~\cite{ls}, the spectrum (in nuclear recoil energy) of
dark matter induced elastic scattering events is given by
\be
\frac{dR}{dE_R} = N_T \frac{\rho_{DM}}{m_{DM}} \int_{|\vec{v}|>v_{\rm
min}} d^3v\, vf(\vec{v},\vec{v_e}) \frac{d\sigma}{d E_R},
\label{rate1}
\ee
where $N_T$ is the number of target nuclei, $m_{DM}$ is the mass of
the dark matter particle, $\rho_{DM}$ is the local dark matter
density, $\vec{v}$ is the dark matter velocity in the frame of the
Earth, $\vec{v_e}$ is the velocity of the Earth with respect 
to the galactic halo, and $f(\vec{v},\vec{v_e})$ is the distribution
function of dark matter particle velocities, which we take to be 
the standard Maxwell-Boltzmann distribution:
\be
f(\vec{v},\vec{v_e}) = \frac{1}{(\pi v_0^2)^{3/2}} {\rm
e}^{-(\vec{v}+\vec{v_e})^2/v_0^2}.
\ee
The Earth's speed relative to the galactic halo is given by
$v_e=v_{\odot}+v_{\rm orb}{\rm cos}\,\gamma\, {\rm
cos}[\omega(t-t_0)]$ where $v_{\odot}=v_0+12\,{\rm km/s}$, 
$v_{\rm orb}=30 {\rm km/s}$, ${\rm cos}\,\gamma=0.51$, $t_0={\rm June
\, 2nd}$, and $\omega=2\pi/{\rm year}$. We take $v_0=230$ km/s and
limit the velocity distribution with a galactic escape velocity of
600 km/s~\cite{ls}. The minimum dark matter 
velocity required to impart a recoil of energy, $E_R$, is given by
$v_{\rm min} = \sqrt{E_R m_N/2 \mu^2}$, where $m_N$ is the mass of
the target nucleus and $\mu$ is the reduced mass of the dark matter
particle and the target nucleus. Throughout our analysis, we take
$\rho_{DM}=0.3$ GeV/cm$^3$.

For a spin-independent cross section between dark matter particles
and nuclei, we have
\be
\frac{d\sigma}{d E_R} = \frac{m_N}{2 v^2} \frac{\sigma_n}{\mu_n^2}
\frac{\left[f_p Z+f_n (A-Z)\right]^2}{f_n^2} F^2(q),
\label{cross1}
\ee
where $\mu_n$ is the reduced mass of the dark matter particle and
nucleon (proton or neutron), 
$\sigma_n$
 is the scattering cross section of the dark matter 
particle with 
neutrons,
$Z$ and $A$ are the atomic and mass numbers of the nucleus, and
$f_{n,p}$ are the coupling strengths of the dark matter particle to
neutrons and protons respectively (we will assume that $f_p=f_n$).
The nuclear form factor, $F(q)$, accounts for the finite momentum
transfer in scattering events.  In our calculations, we adopt the
Helm form factor:
\begin{equation}
F(q) = \frac{3 j_1(q R_1)}{q R_1} \,\, e^{-{\frac{1}{2}q^2 s^2}}, 
\end{equation}
where $j_1$ is the second spherical bessel function and $R_1$ is
given by
\begin{equation}
R_1 = \sqrt{c^2 + \frac{7 \pi^2 a^2}{3} - 5 s^2}.
\end{equation}
Here, $c \approx 1.23 A^{1/3} -0.60$ fm, $a \approx 0.523$ fm, and $s
\approx 0.9$ fm have been determined by fits to nuclear physics
data~\cite{Gondolo,Fricke}. Note that other commonly used
parameterizations of the form factor can lead to modest but not
insignificant (on the order of 10 to 20\%) variations in the region
of dark matter parameter space that provide a good fit to the CoGeNT
(and to a lesser extent DAMA/LIBRA) signal.  

While variations in the velocity distribution of dark matter
particles could also significantly affect the quality of the fits
found to the CoGeNT and/or DAMA/LIBRA data (see, for example,
Ref.~\cite{McCabe:2010zh}), such changes tend to affect the fits to
each data set in a similar way. Increasing $v_0$ and/or $v_{\rm
esc}$, for example, will tend to move the acceptable regions of dark
matter parameter space toward lighter masses (and smaller cross
sections) for both CoGeNT and DAMA/LIBRA. Since both regions will be
moved in approximate unison, we do not consider such variations
further. Similarly, we do not contemplate any deviations from a standard isothermal dark matter halo, another source of possible uncertainty affecting the comparison of DAMA/LIBRA 
to other experiments~\cite{cowsik}.

\begin{figure}[t]
\centering
{\includegraphics[angle=0.0,width=3.5in, trim= 40 170 190 300, clip=true]{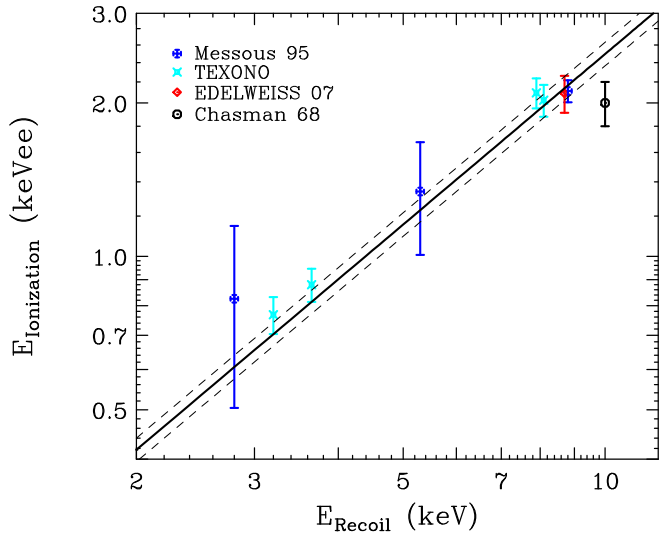}}
\caption{Measurements of the germanium quenching factor ($Q_{\rm Ge}
\equiv E_{\rm ionization}/E_{\rm Recoil}$) over the energy range of
the excess events observed by CoGeNT. The solid line denotes the best
fit normalization to these measurements, assuming the slope predicted
by Lindhard theory ($k=0.20$). The dashed lines represent the upper
and lower $2 \sigma$ normalizations, accounting only for statistical
errors. For the measurements used, see Ref.~\cite{Ge}. Additional 
measurements by the CoGeNT collaboration span down to $E_{\rm Recoil}=0.7$ keV~\cite{Ge2}.}
\label{gefit}
\end{figure}

Over the energy range of the CoGeNT signal (approximately 0.4 to 2
keVee, where keVee denotes the equivalent electron energy), a number
of measurements have been made of the relevant quenching factors
({\it i.e.}~the ratio of ionization energy to total recoil
energy)~\cite{Ge,Ge2}. These are summarized in Fig.~\ref{gefit}. The
solid line in this figure represents the best fit to the data shown,
assuming a parametrization chosen to follow the Lindhard
theory (using $k=0.20$). The dashed lines reflect the 2$\sigma$
statistical upper and lower limits. In our fits, we will adopt a
quenching factor for germanium given by $Q_{\rm Ge}(E_{\rm
Recoil}=3\,{\rm keV})=0.218 \pm 0.0058$, and with the energy
dependence predicted by the Lindhard theory. Note that this neglects
any systematic errors; the inclusion of which would further enlarge
the region of dark matter parameter space potentially capable of
accommodating the CoGeNT signal.


\begin{figure}[t]
\centering
\includegraphics[width=1.0\columnwidth]{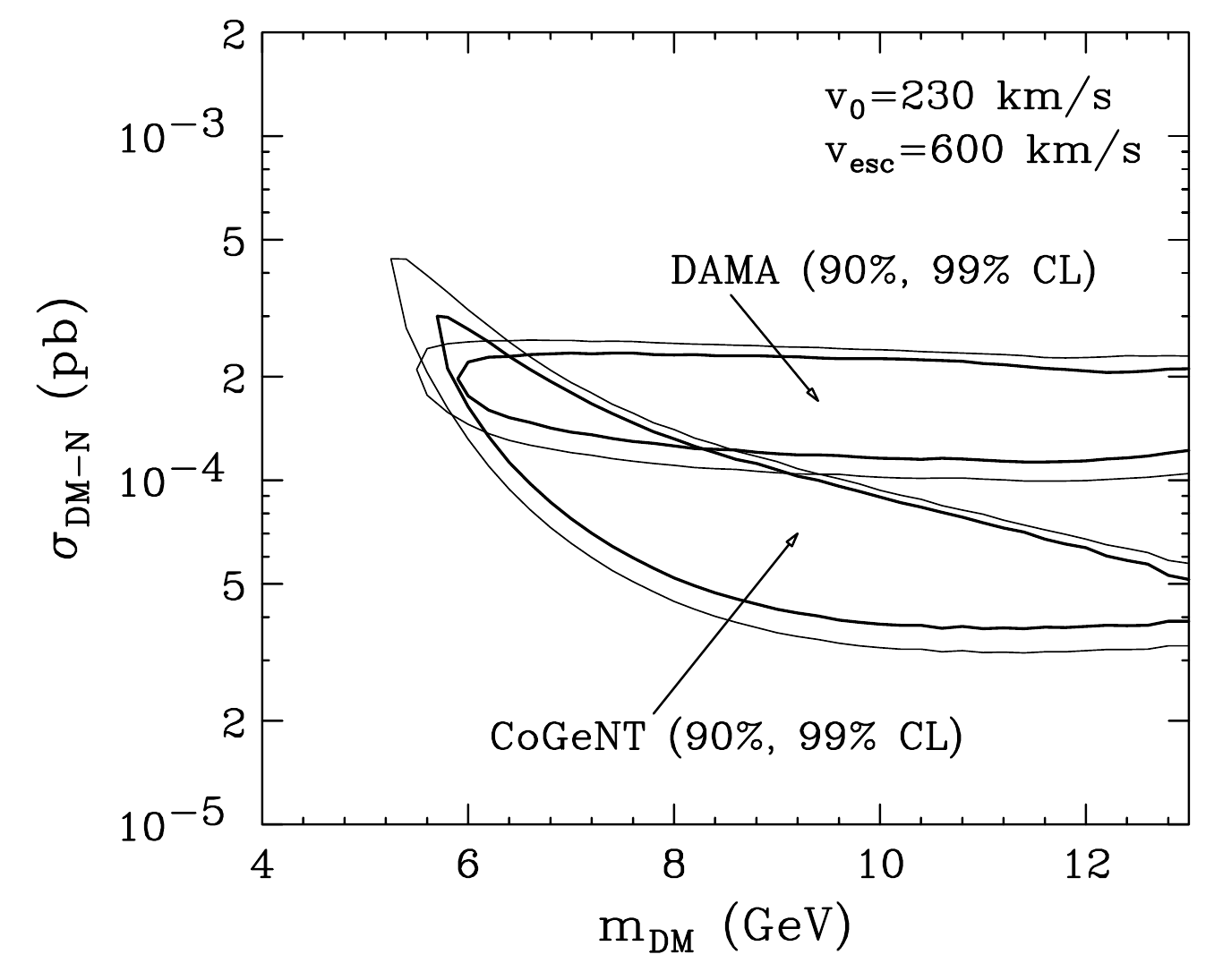}
\includegraphics[width=1.02\columnwidth]{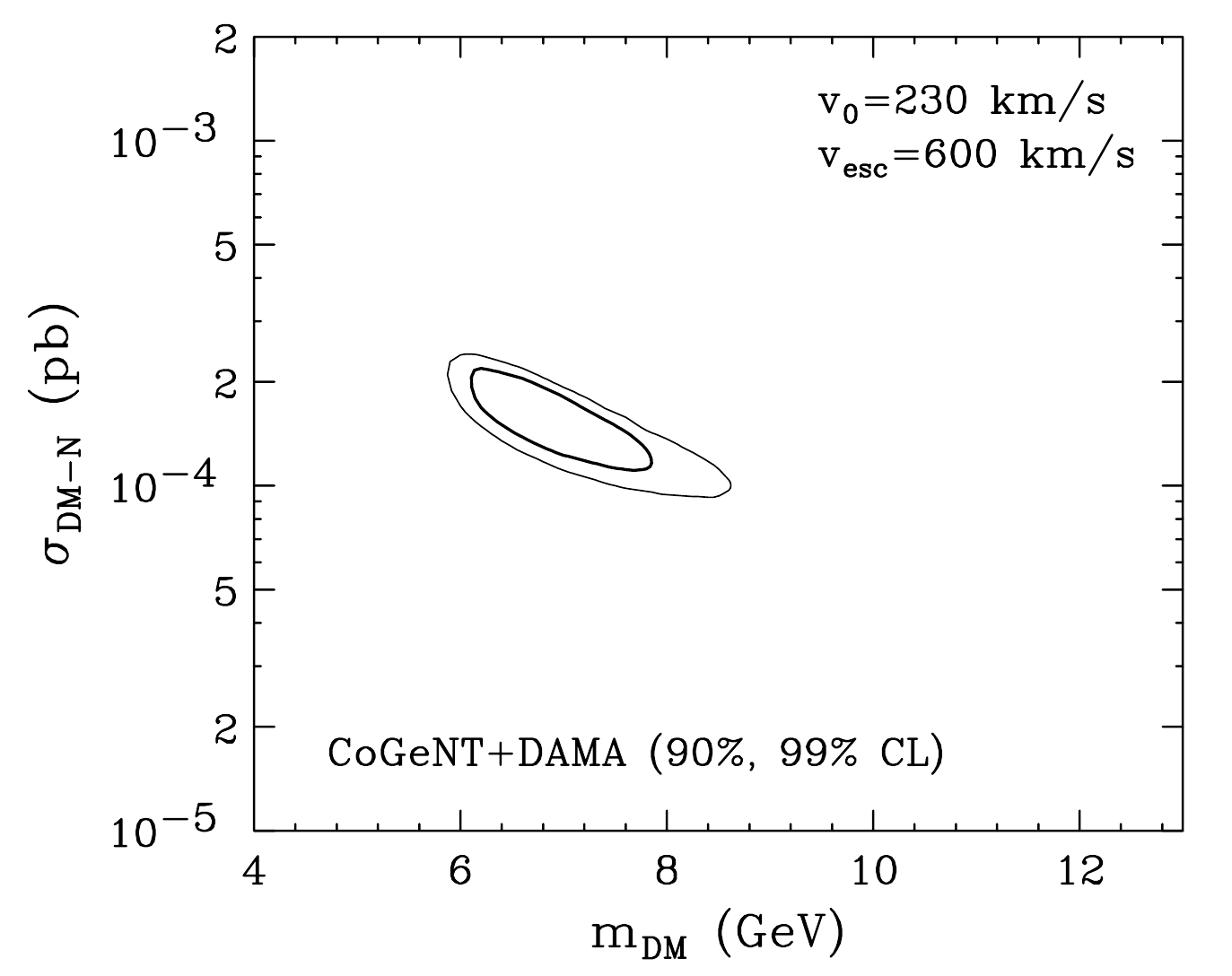}
\caption{The regions in the elastic scattering cross section (per
nucleon), mass plane in which dark matter provides a good fit to the
excess CoGeNT events and to the annual modulation reported by
DAMA/LIBRA (upper frame), as well as the region in which the
combination of CoGeNT+DAMA/LIBRA is well fit (lower frame). We have
assumed that any effects of channeling are negligible and have
adopted $v_0=230$ km/s and $v_{\rm esc}=600$ km/s. No errors
associated with uncertainties in the form factors have been taken
into account. If these and other systematics were fully included, the
allowed region would be expected to increase considerably. See text
for more details.}
\label{fits}
\end{figure}

For DAMA/LIBRA, measurements of the NaI(Tl) quenching factors are often
averaged over large ranges of energy, hindering efforts to quantify
the uncertainties in the narrow energy range of interest for light
dark matter particles. In particular, the DAMA/LIBRA collaboration
reports a measurement of their sodium (in the form of NaI, doped with
thallium) quenching factor to be $Q_{\rm Na} = 0.30 \pm 0.01$
averaged over the energy recoil range of 6.5 to 97
keV~\cite{damaquenching}. Other groups have reported similar values:
$Q_{\rm Na} = 0.25 \pm 0.03$ (over 20-80 keV), $0.275 \pm 0.018$
(over 4-252 keV), and $0.4 \pm 0.2$ (over 5-100
keV)~\cite{Fushimi:1993nq}. As the sodium quenching factor is
generally anticipated to vary as a function of energy, it is very
plausible that over the range of recoil energies relevant for light
(5-10) GeV dark matter (approximately 5 to 20 keV) the quenching
factor could be somewhat higher than the average values reported from
these measurements~\cite{fc} (see, for example,
Ref.~\cite{Tretyak:2009sr} and discussion in Ref.~\cite{leffuncertain}). For recoil energies below approximately
20 keV, Ref.~\cite{Tovey:1998ex} reports a measurement of $Q_{\rm Na}
= 0.33 \pm 0.15$, whereas Ref.~\cite{spooner} reports a somewhat
smaller value of $Q_{\rm Na}=0.252 \pm 0.064$ near 10 keV.  
A failure to account for the non-proportionality in electron 
response at low energy~\cite{nonprop} appears in the energy 
calibration of several of these 
measurements, including those of Ref.~\cite{spooner}: the need for additional 
precision measurements of quenching factor 
near DAMA/LIBRA's threshold of 2 keVee 
seems evident. In our
fits, we conservatively adopt a sodium quenching factor of $Q_{\rm
Na}=0.3 \pm 0.13 $ over the energy range of interest ($E \approx
2-6$ keVee), which we deem representative of present experimental 
uncertainties.

\begin{figure}[t]
\centering
\includegraphics[width=1.0\columnwidth]{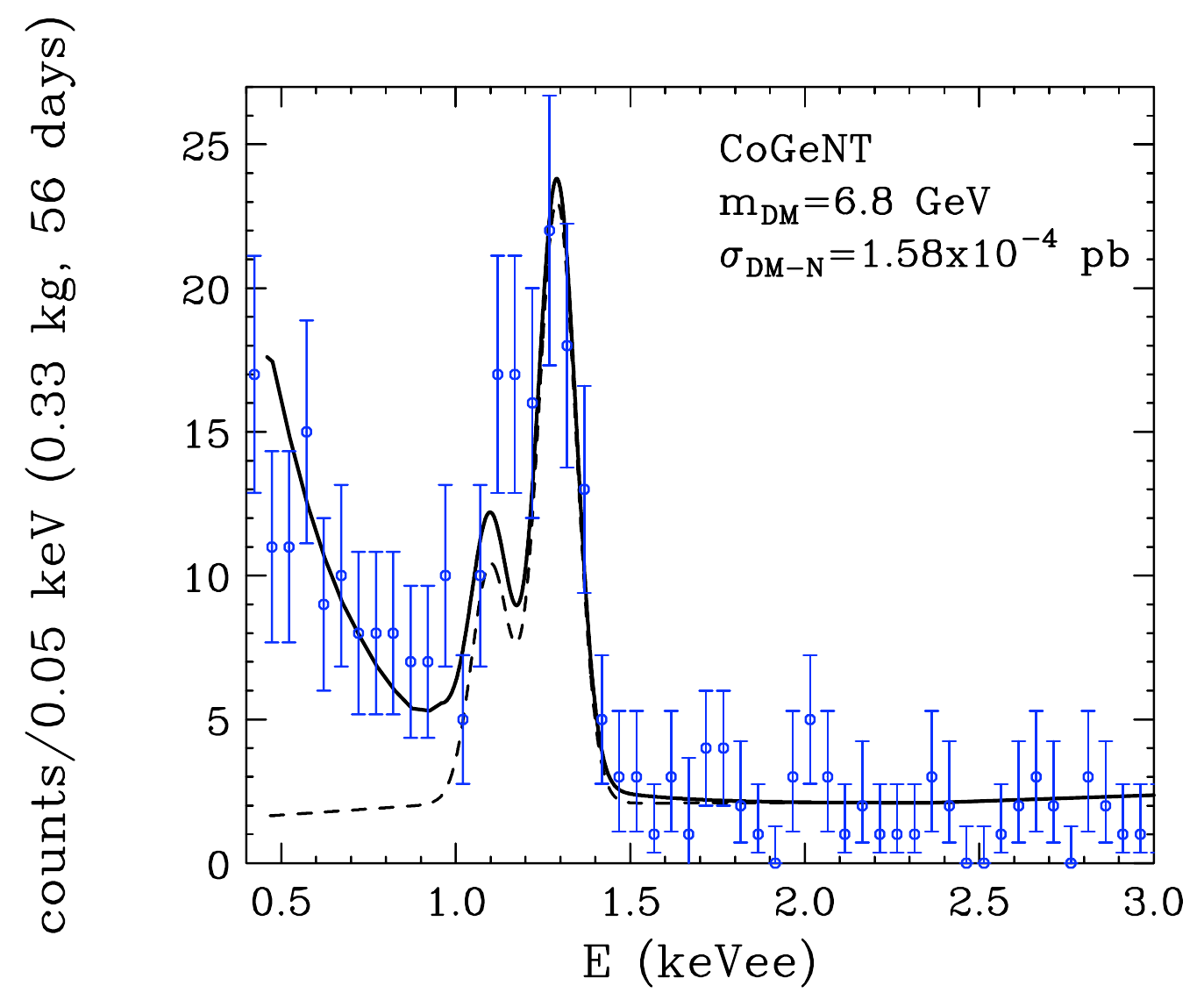}
\includegraphics[width=1.0\columnwidth]{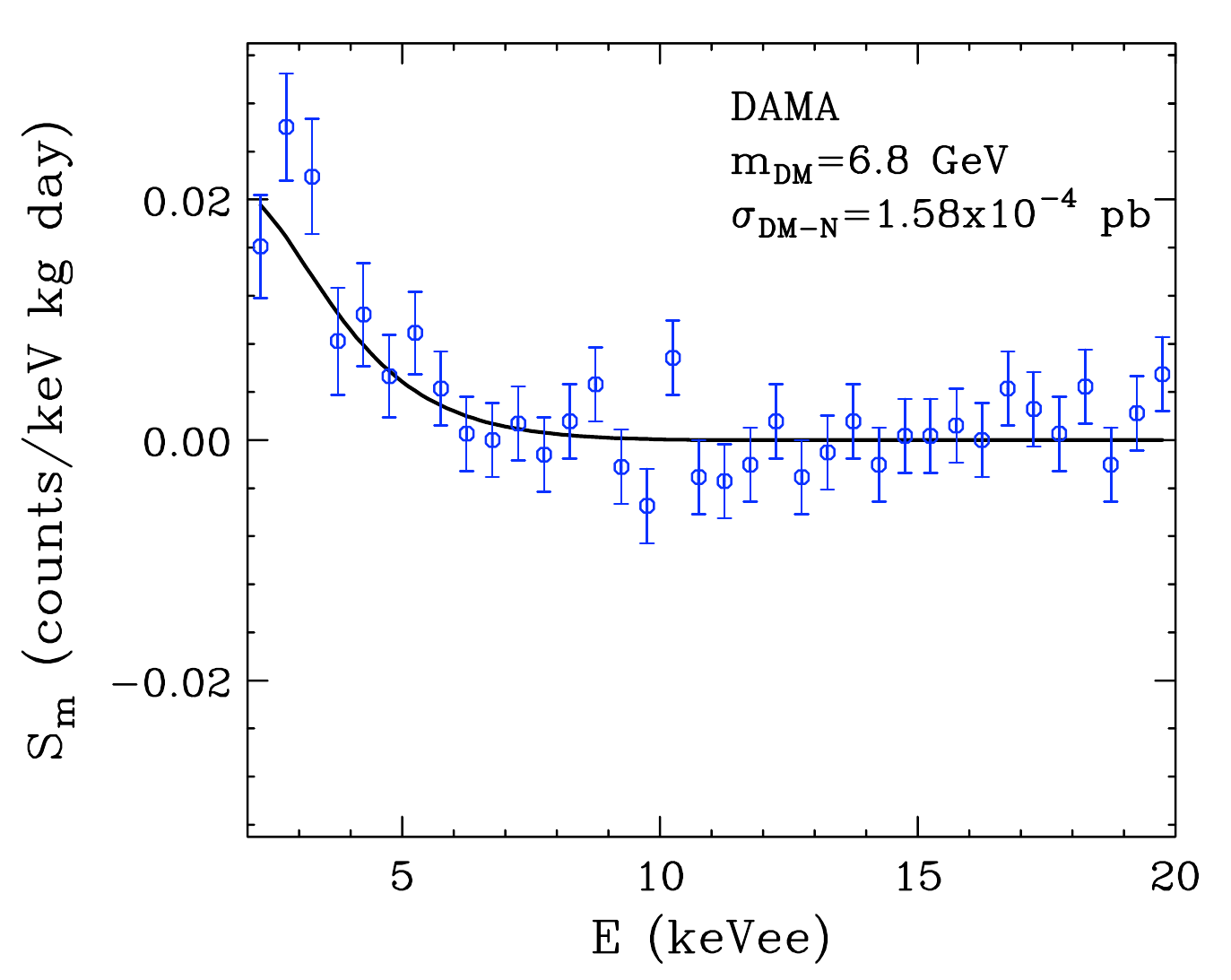}
\caption{The spectrum of events in CoGeNT (upper frame) and the
spectrum of the annual modulation in DAMA/LIBRA (lower frame) for
overall best fit dark matter parameters of $m_{\rm DM}=6.8$ GeV and
$\sigma_{\rm DM-N}=1.58\times 10^{-4}$ pb. In the upper frame, the
solid black line is the predicted result for signal plus
background (with triggering and signal acceptance efficiency built 
into the model), whereas the dashed line is the background alone and
points denote the measured values. In the lower
frame, the solid line is the predicted signal and the points
denote the measurements reported by DAMA/LIBRA. We have assumed that
any effects of channeling are negligible and have adopted $v_0=230$
km/s and $v_{\rm esc}=600$ km/s. See text for more details.}
\label{spec}
\end{figure}

In Fig.~\ref{fits}, we show the regions of dark matter parameter
space which provide a good fit to the DAMA/LIBRA and CoGeNT data
separately (upper frame) and combined (lower frame). In performing
our fits, we have used the (13) DAMA/LIBRA bins below 8.5 keVee and
the (28) CoGeNT bins between 0.4 and 1.8 keVee. The data at higher
energies will not include any events from dark matter particles in
the mass range considered here, and the inclusion of higher energy
bins would not affect our results in any significant way. 

From Fig.~\ref{fits}, we see that there exists a range of masses and
cross sections for which both DAMA/LIBRA and CoGeNT can potentially
be accommodated. In the range of $m_{\rm DM} \sim$7-8 GeV and
$\sigma_{\rm DM-N}\approx (1-3) \times 10^{-4}$ pb, quite good fits
can be found for both experiments.\footnote{An eventual stripping of L-shell 
electron capture 
peaks in the low-energy CoGeNT spectrum, based on high-statistics 
measurements of their K-shell counterparts and the known L/K capture 
ratio~\cite{Bahcall}, is 
expected to favor precisely this same dark matter mass and cross section.} The overlapping region requires fairly large values of the sodium quenching factors, $Q_{\rm Na} \approx 0.45$ or greater throughout the 99\% CL region and $Q_{\rm Na} \approx 0.50-0.55$ in the 90\% CL region; considerably larger than the measurements presented in Ref.~\cite{spooner}. In the upper frame of Fig.~\ref{spec}, we show the spectrum of events in CoGeNT for the
case of $m_{\rm DM}=6.8$ GeV and $\sigma_{\rm DM-N}=1.58 \times
10^{-4}$ pb. The dashed line shows our background model, which
consists of a flat spectrum combined with a well understood double
gaussian peak (see Ref.~\cite{cogentnew} for details). In the lower
frame of Fig.~\ref{spec}, we show the prediction for the same dark
matter model compared to the spectrum of DAMA/LIBRA's annual
modulation. From these plots, it is clear that both the CoGeNT and
DAMA/LIBRA signals could potentially result from a $\sim$7-8 GeV dark
matter particle with an elastic scattering cross section of
$\sigma_{\rm DM-N}\approx (1-3) \times 10^{-4}$ pb.

\begin{figure}[t]
\centering
\includegraphics[width=1.0\columnwidth]{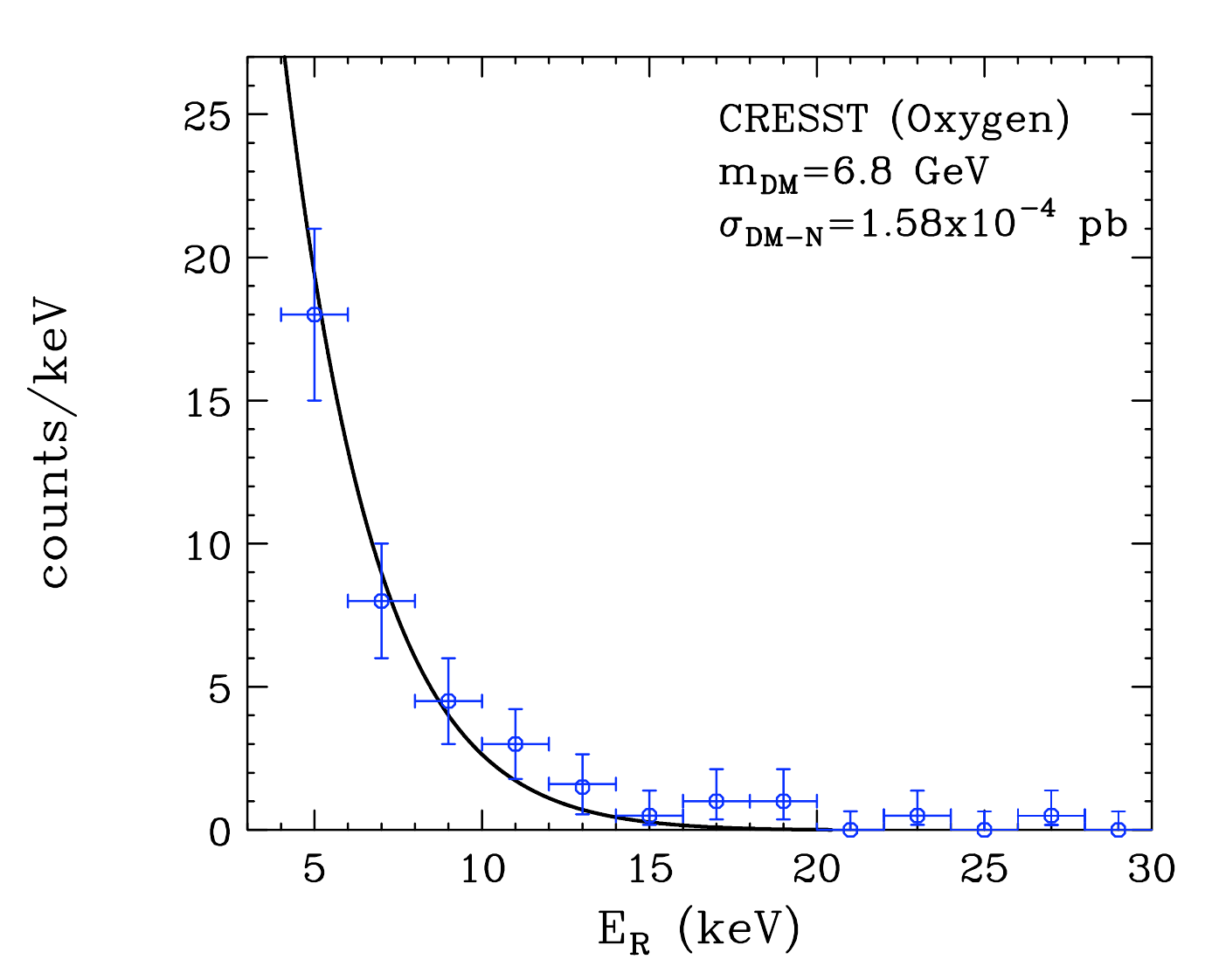}
\caption{The preliminary spectrum of events in the oxygen band of the
CRESST experiment, compared to the spectral shape predicted for the
case of $m_{\rm DM}=6.8$ GeV and $\sigma_{\rm DM-N}=1.58\times
10^{-4}$ pb (which provides good fit to both CoGeNT and DAMA/LIBRA).
The solid line is the predicted signal and the error bars denote the
preliminary spectrum of events reported by the CRESST collaboration.
We have adopted $v_0=230$ km/s and $v_{\rm esc}=600$ km/s. See text
for more details.}
\label{speccresst}
\end{figure}

Lastly, we briefly consider the spectrum of events reported in recent
talks by the CRESST collaboration~\cite{wonder}. In the
data from 9 CaWO$_4$ crystals, with a total exposure of 333 kg-days,
a larger than anticipated number of events has been observed in the
oxygen band of their experiment with recoil energies below $\sim$20
keV. In Fig.~\ref{speccresst}, we show the spectrum of the oxygen
band events reported in Ref.~\cite{wonder} and compare this to the
spectrum predicted for a dark matter particle consistent with both
CoGeNT and DAMA/LIBRA ($m_{\rm}=6.8$ GeV, $\sigma_{\rm DM-N}=2\times
10^{-4}$ pb). Note that as the total exposure of the observation is
not completely specified in Ref.~\cite{wonder}, we have normalized the predicted
curve (the solid line) to the data, which corresponds to an exposure
(times efficiency) of 210 kg-days. Remarkably good agreement is
found. For heavier dark matter particles, most of the dark matter
events are expected to result from scattering with tungsten rather
than oxygen nuclei. In the case of a very light dark matter particle,
however, scattering with tungsten produces events with recoil
energies below the threshold of the experiment. For this dark matter
mass and cross section, we predict only one event in the tungsten
band above 3.7 keV, and about ten events between 3.0 and 3.7 keV. 
We would like to emphasize the preliminary nature of these results,
and recognize that, until the CRESST collaboration publishes their
final distribution of events, fits to these data should be assessed with caution. In particular, we emphasize that some
fraction of the events observed in the oxygen band could be spillage
from CRESST's alpha or tungsten bands, neutron backgrounds, or be the
result of radioactive backgrounds. Further information from the CRESST collaboration will be essential for understanding these results.


\section{Consistency With Null Results}
\label{null}

\begin{figure}[t]
\centering
\includegraphics[width=1.0\columnwidth]{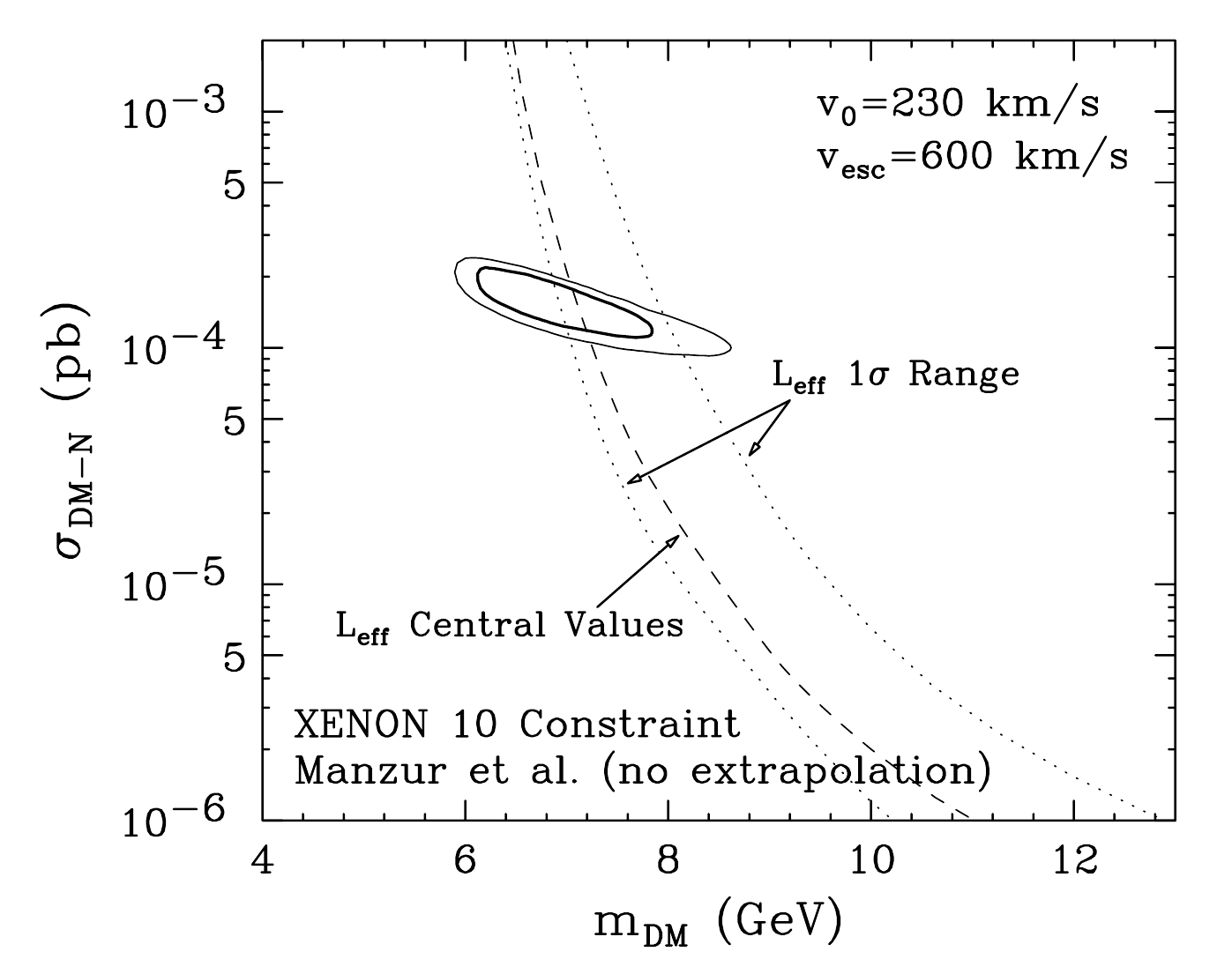}
\includegraphics[width=1.01\columnwidth]{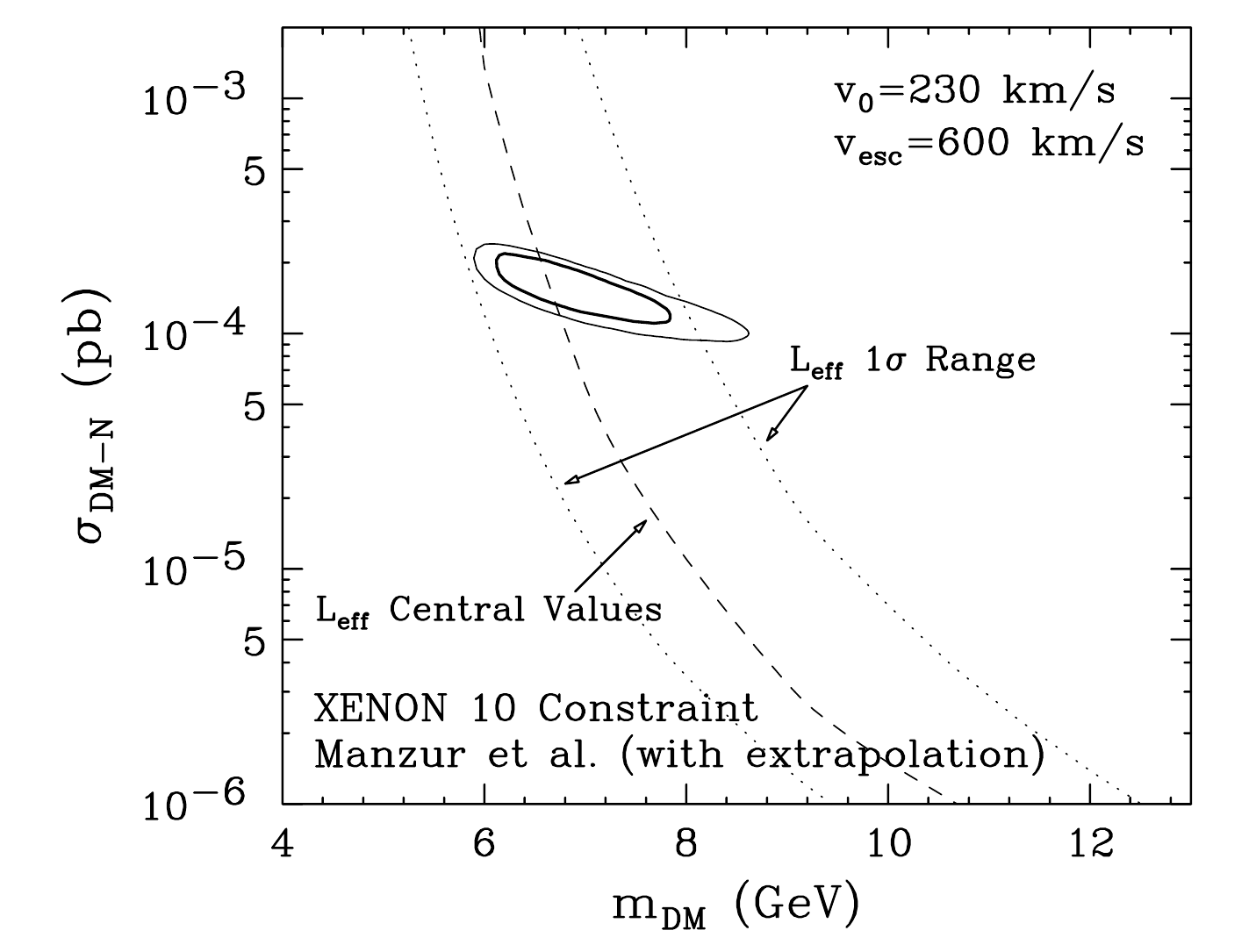}
\caption{Constraints from the XENON 10 experiment~\cite{Savage:2010tg}. 
In each frame, the
dashed line denotes the limit when using the central values of the
scintillation efficiency, $L_{\rm eff}$, as measured by Manzur {\it
et al.}~\cite{manzur}, whereas the dotted lines are derived using
$\pm 1 \sigma$ values of $L_{\rm eff}$. In the upper frame, no
assumptions are made regarding the values of $L_{\rm eff}$ at
energies below 4 keV (for which no measurements exist). In the lower
frame, $L_{\rm eff}$ is assumed to fall linearly below 4 keV. 
Considerably more relaxed constraints are obtained from other 
existing measurements of $L_{\rm eff}$~\cite{leffuncertain}. 
See text for more details.}
\label{xenon}
\end{figure}

In this section, we discuss whether a dark matter interpretation of
the combined CoGeNT and DAMA/LIBRA signals is consistent with the
null results reported by other direct detection experiments. In
particular, recent claims have been made that a dark matter
interpretation of the CoGeNT and DAMA/LIBRA data is inconsistent with the
measurements of the XENON 100 experiment~\cite{xenon100}. This
conclusion, however, depends critically on the scintillation
efficiency of liquid xenon, $L_{\rm eff}$ that is
adopted~\cite{leffuncertain,Savage:2010tg}. In particular, while both
theoretical arguments and measurements of $L_{\rm eff}$ lead one to
expect this quantity to decrease at low energies, no measurements
exist below $\sim 4$ keV, forcing one to speculate or extrapolate at
lower energies. Unless quite optimistic values for these quantities
are adopted, the range of masses and cross sections best fit by
DAMA/LIBRA and CoGeNT are not significantly constrained by XENON 100~\cite{leffuncertain}.
In fact, stronger constraints than those from XENON 100 can be
derived from the data of XENON 10, due to its lower energy
threshold~\cite{Savage:2010tg} (see also Ref.~\cite{xenonnull}).
The recent work of Manzur {it et al.} provides measurements of $L_{\rm eff}$ over
the range of approximately 4 to 70 keV~\cite{manzur}. By not taking into account Poisson
fluctuations from dark matter signals below 4 keV, and thus not
making any assumptions regarding the values of $L_{\rm eff}$ below this
range, it is possible to arrive at the constraints shown in the upper frame of
Fig.~\ref{xenon}. These constraints yield only a mild tension (less
than $\sim$1$\sigma$) with the parameter space region favored by
DAMA/LIBRA and CoGeNT. If we instead assume that $L_{\rm eff}$ drops
linearly below 4 keV, slightly stronger limits are found
(Fig.~\ref{xenon}, lower frame). Again, however, this constraint
conflicts with the region favored by DAMA/LIBRA and CoGeNT at only
about $\sim$1$\sigma$. We emphasize that other existing measurements and 
extrapolations of 
$L_{\rm eff}$ lead to a complete absence of constraints on the region of 
DAMA/LIBRA and CoGeNT compatibility, even when sub-threshold 
Poisson fluctuations are 
assumed~\cite{leffuncertain}.

\begin{figure}[t]
\centering
\includegraphics[width=1.0\columnwidth]{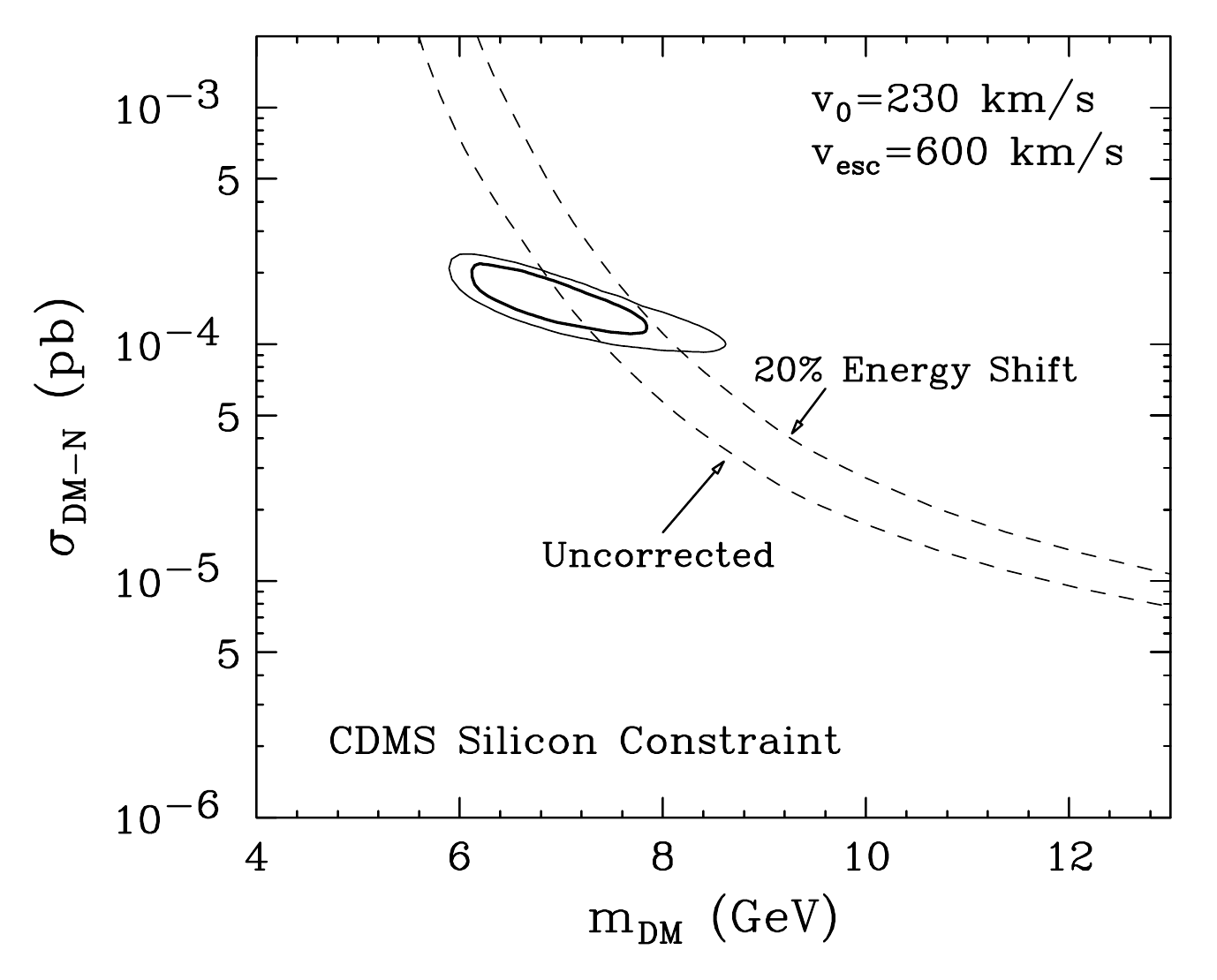}
\caption{Constraints from the CDMS experiment's silicon analysis. The
lower dashed curve denotes the results as presented in
Ref.~\cite{filippini,filippini2}, whereas the upper dashed curve
shows the result with a 20\% shift in CDMS's silicon recoil energy
scale, a conservative correction that alleviates concerns expressed 
in the discussion surrounding Fig.~3.20 of Ref.~\cite{filippini}.
See text for more details.}
\label{cdmssi}
\end{figure}

For typical dark matter masses, the null results from CDMS-II's
germanium detectors provide the strongest constraints on the dark
matter-nucleon elastic scattering cross-section~\cite{cdms}.  Below
$\sim$10 GeV, however, the CDMS-II silicon detectors provide
better constraints~\cite{filippini,filippini2} due to the favorable
kinematics of the lighter target nucleus.  In Fig.~\ref{cdmssi}, we
compare these contraints to the regions favored by the dark matter
interpretation of the combined DAMA/LIBRA and CoGeNT results.  Taken
as published (after accounting for the different velocity
distribution used in Refs.~\cite{filippini,filippini2}), we find that
this constraint covers most of the $2\sigma$ range of masses and
cross sections found to fit the DAMA/LIBRA and CoGeNT signals. 

As noted in Ref.~\cite{filippini} (and as shown in their Fig.~3.20),
however, the observed CDMS-II silicon nuclear recoil quenching is not
reproduced by Lindhard theory, and is also markedly discrepant with previous
measurements~\cite{dougherty}. In contrast, an excellent agreement is 
observed
for CDMS germanium detectors.\footnote{We remark without undue 
emphasis that a rough analysis of the CDMS germanium data in the relevant 2-5 keV recoil energy region exists~\cite{ogburn}.} It is possible to
attribute this disagreement to a systematic error in the absolute
energy scale in the silicon detectors. The energy scale of the
silicon detectors is more complicated than the germanium detectors to
calibrate, since the silicon detectors are not thick enough to
contain the full energy deposition from barium gamma rays used for 
calibration. Additionally, large corrections affecting the recoil 
energy scale are applied to the CDMS detectors to remove position dependances (see the discussion surrounding Fig.~3.18 of Ref.~\cite{filippini}).
The discrepancy between the observed quenching and Lindhard theory could indicate a
$\sim$20-30\% error in the low energy calibration, larger if other
existing experimental data~\cite{dougherty} are taken as the reference.  
In Fig.~\ref{cdmssi},
we show how a corrected energy scale can change the constraints derived from
the CDMS-II experiment, for the case of a linear $20\%$
correction.\footnote{Note that a non-linear energy correction would
be needed to reconcile the Lindhard theory with the energies observed
at CDMS-II. In particular, Fig~3.20 of Ref.~\cite{filippini} shows
the observed nuclear recoil band crossing the prediction from
Lindhard theory. The linear 20\% correction used here, however,
represents a reasonable estimate for the range of energies relevant
for the detection of $\lsim 10$ GeV dark matter.} This shows that,
while the CDMS-II silicon exposure could potentially constrain the
region favored by DAMA/LIBRA and CoGeNT, this constraint is weakened
due to the energy scale uncertainty and does not rule out the region
favored by these experiments.



\section{Summary and Discussion}
\label{conclusions}

In this paper, we have studied the excess of low energy events
recently reported by the CoGeNT collaboration and the annual
modulation signal reported by DAMA/LIBRA and conclude that these two
signals could arise from an elastically scattering dark matter
particle with a mass in the approximate range of $\sim$7 GeV and a
cross section (with nucleons) of $\sigma \sim 2 \times 10^{-4}$ pb
($2 \times 10^{-40}$ cm$^2$). This conclusion is reached even if
channeling is assumed to be negligible. The concordance between these
two signals, which has not been found in previous studies, is made
possible in large part by our choice of nuclear form factors and our
accounting for uncertainties in the quenching factors of germanium
and sodium. We also point out that the preliminary events observed in the oxygen
band of the CRESST experiment are consistent with being the result of
such a dark matter particle. 

We have also considered in this paper the constraints from null
results of other direct detection experiments, including XENON 10,
XENON 100, and CDMS (Si). After taking into account the uncertainties
in the scintillation efficiency of liquid xenon and the recoil energy
scale of silicon events at CDMS, we find that the region of dark
matter parameter space favored by CoGeNT and DAMA/LIBRA is consistent
with all current constraints.

In the future, it may become possible for the CoGeNT or CRESST
experiments to observe an annual modulation in their rate. In
particular, we calculate that if CoGeNT is observing dark matter 
interactions, their
event rate should be approximately 20\% higher in the summer than it
is in the winter, for a particle of this mass and CoGeNT's energy 
threshold. To detect this effect with a significance of
3$\sigma$, an approximate exposure of 40 kg-days would be required in
each of the summer and winter seasons. This goal appears to be
attainable for the CoGeNT experiment, which is operating continuously 
since December of 2009 with an active target mass of 0.33 kg. 
If observed, this would provide an important confirmation of the hypothesis that these experiments are in fact detecting dark matter. 


\section*{Acknowledgements} We would like to thank Fabio Cappella,
Chris Savage, Kathryn Freese, Jeff Filippini, Peter Sorensen, 
Vladimir Tretyak, Neal Weiner, and Kathryn Zurek for helpful discussions. DH would also like to thank the Aspen Center for Physics, where some of this work was completed. DH is supported by
the US Department of Energy, including grant DE-FG02-95ER40896, and
by NASA grant NAG5-10842.

\end{document}